\documentclass{ws-p8-50x6-00}

\usepackage{floatflt}

\begin{document}

\title{Recent NA49 results on Pb+Pb collisions\\at CERN SPS}

\author{Ferenc Sikl\'er for the NA49 Collaboration}

\address{KFKI Research Institute for Particle and Nuclear Physics, Budapest, Hungary\\
E-mail: sikler@rmki.kfki.hu}  

\maketitle


\abstracts{In the spirit of establishing a fair reference for
nucleus-nucleus collisions, results on stopping and baryon transfer,
correlations of the p+p interaction and their consequences are shown. In
the discussion of new results from nucleus-nucleus collisions the emphasis
is on strange meson and baryon production at different energies -- for the
first time at 40 GeV$\cdot A$ -- with the study of light nuclei.}

\section{Motivation}

We study nuclear reactions in order to elucidate the strong interaction
process and study its development in time and space with the hope to learn
the characteristics of hot and dense matter, moreover to establish
experimental links between elementary and complex interactions.

\section{The experiment}

The NA49 experiment is a large acceptance detector for charged
hadrons\cite{nim}. The tracking of particles is based on four large volume
time projection chambers, the particle identification is possible via the
measurement of their specific energy loss ($dE/dx$).

One of the key parameters in the systematics of hadron production is the
centrality of the collision: the number of collisions per participant
nucleon. In case of p+A reactions the centrality is known to be correlated
with the number of "grey" particles -- slow protons and deuterons -- which
are measured by a centrality detector surrounding the target and by the
tracking system. In Pb+Pb reactions the deposited energy in the zero
degree calorimeter was used to select events with given centrality. Both
measures were then correlated with the average number of collisions a
nucleon undergoes, using simulation.

The data presented here are preliminary results from p+p, p+Al, p+Pb and
C+C, Si+Si, Pb+Pb reactions at 158 GeV$\cdot A$ and from Pb+Pb reactions
at 40 GeV$\cdot A$ energy.

\section{Stopping and baryon transfer}

\begin{figure}[h]
\epsfxsize160pt
\figurebox{}{}{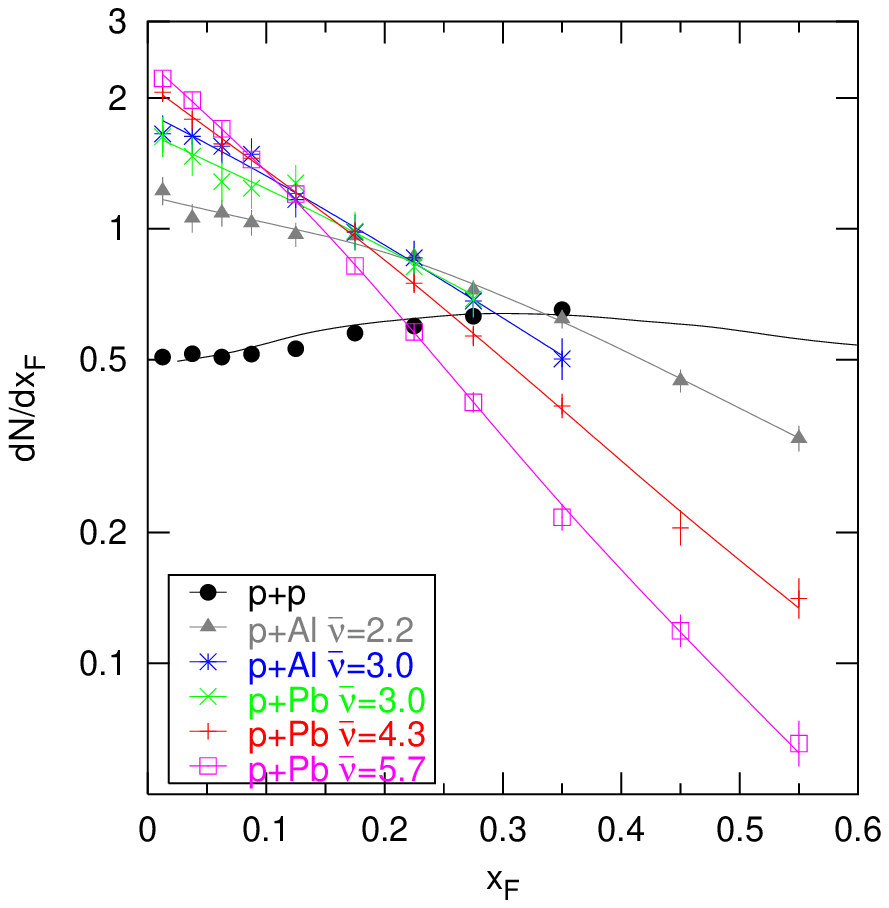}
\figurebox{}{}{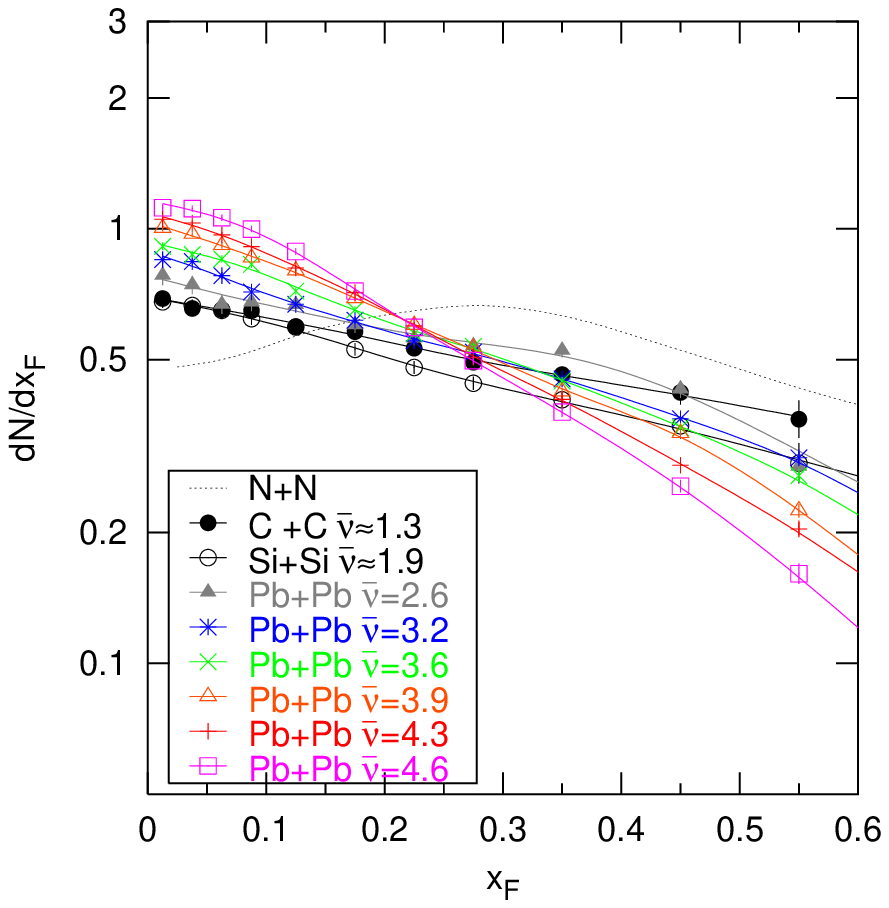}

\caption{Feynman-x distribution of net protons ($\mathrm{p-\bar{p}}$) with
different average number of collisions ($\bar{\nu}$) in {\it a)} p+A and
{\it b)} A+A collisions. Lines are to guide the eye.}

\label{fig:net_proton}
\end{figure}

The longitudinal momentum distribution of net protons from
non-single-diffractive p+p interactions has been measured in the rest
frame of the collision (Feynman-x distribution, $x_{_F}$). This can be
compared to the measurements of the net proton distribution in centrality
selected p+A reactions (Fig.~\ref{fig:net_proton}.a). With increasing
centrality (increasing average number of collisions) the final state
protons appear to be more and more stopped with the biggest effect in
the most central sample. The evolution is smooth, connecting p+p via p+Al
to p+Pb.\footnote{Note the similarity of p+Al and p+Pb in the overlap
region at $\bar{\nu}\approx 3$.}

In A+A collisions a similar gradual evolution of stopping is seen
(Fig.~\ref{fig:net_proton}.b), although not reaching the high level
observed in most central p+Pb.

\section{Correlations in p+p, predictions}

A very characteristic systematics of the net proton distribution as
function of $\bar{\nu}$ is observed in p+A and A+A reactions. What about
p+p collisions? Although here $\nu=1$ by definition, the degree of
inelasticity of the interaction can be characterized by the $x_{_F}$ of
the final state proton. This is exemplified in the following by inspecting
the correlations of hadronic variables with $x_{_F}^p$.

 \subsection{Pion density}

\begin{figure}[h]
\epsfxsize160pt
\figurebox{}{}{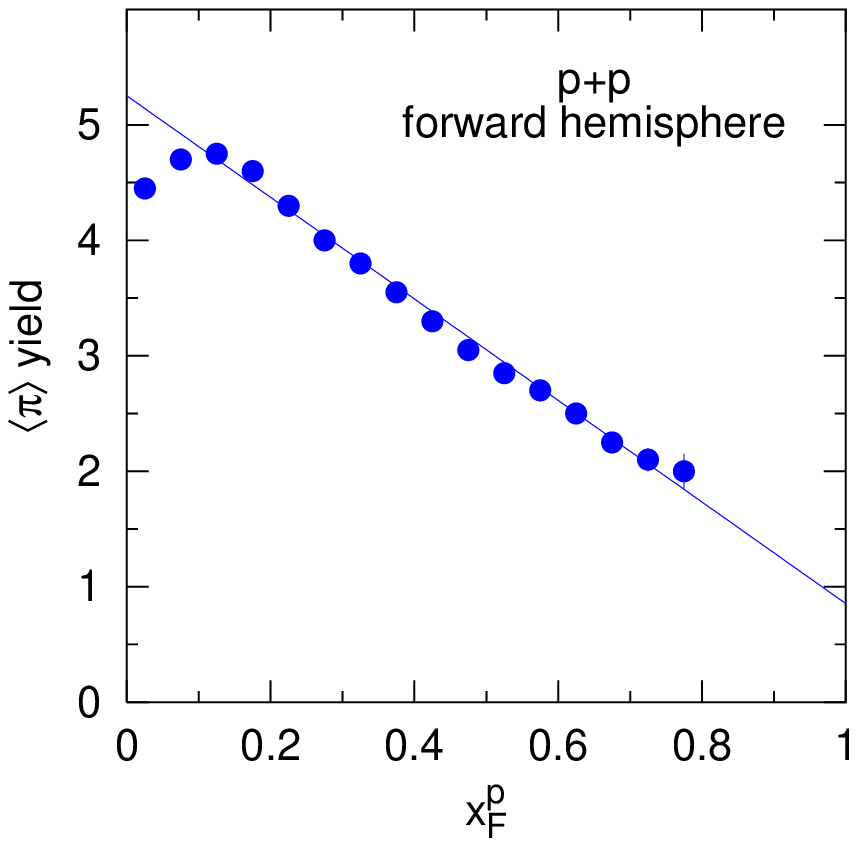}
\figurebox{}{}{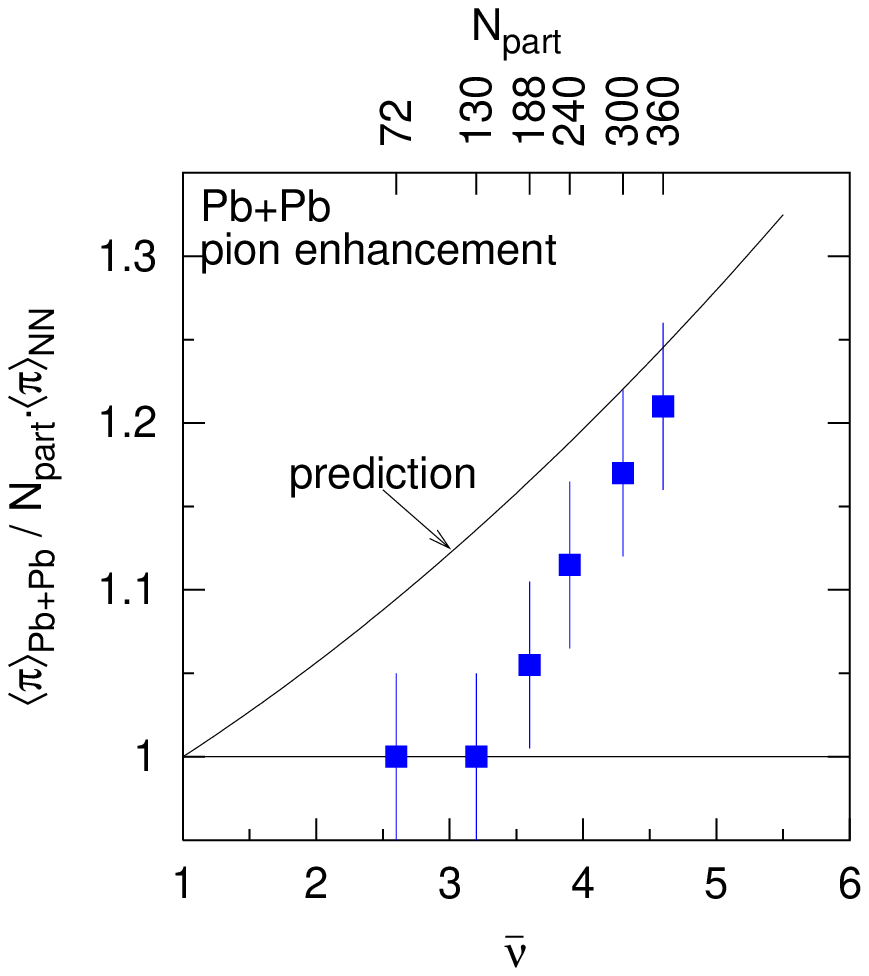}

\caption{{\it a)} Average number of charged pions
($\langle\pi\rangle=(\pi^++\pi^-)/2$) in the forward hemisphere in p+p
collision if the fastest proton has longitudinal momentum $x_{_F}^p$. {\it
b)} Pion enhancement in Pb+Pb collisions relative to minimum bias p+p as
function of $\bar{\nu}$. Measured points and predicted line using p+p
correlations are shown}

\label{fig:pred_pion}
\end{figure}

The average number of charged pions $\langle\pi\rangle$ strongly
correlates with the $x_{_F}$ of the final state proton
(Fig.~\ref{fig:pred_pion}.a): a fast proton will be accompanied by a few,
a slow one by many pions. 

Using the $\langle\pi\rangle$ -- $x_{_F}^p$ correlation measured in p+p,
predictions can be made for other reactions by folding their -- above
discussed -- proton distribution with this correlation curve. For a
reaction that has slower protons one would extrapolate bigger pion
density. This is what happens in more and more central Pb+Pb collisions:
the observed increase of pion density with centrality is close to the
prediction from p+p interactions (Fig.~\ref{fig:pred_pion}.b).

 \subsection{Strangeness content}

\begin{figure}[h]
\epsfxsize160pt
\figurebox{}{}{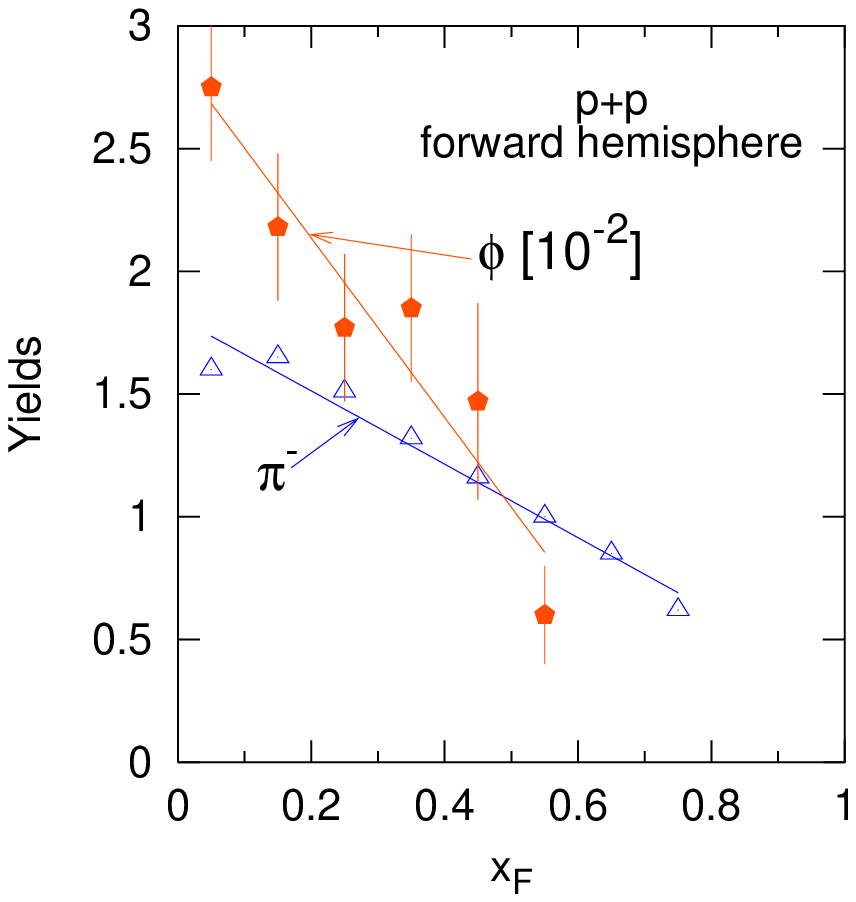}
\figurebox{}{}{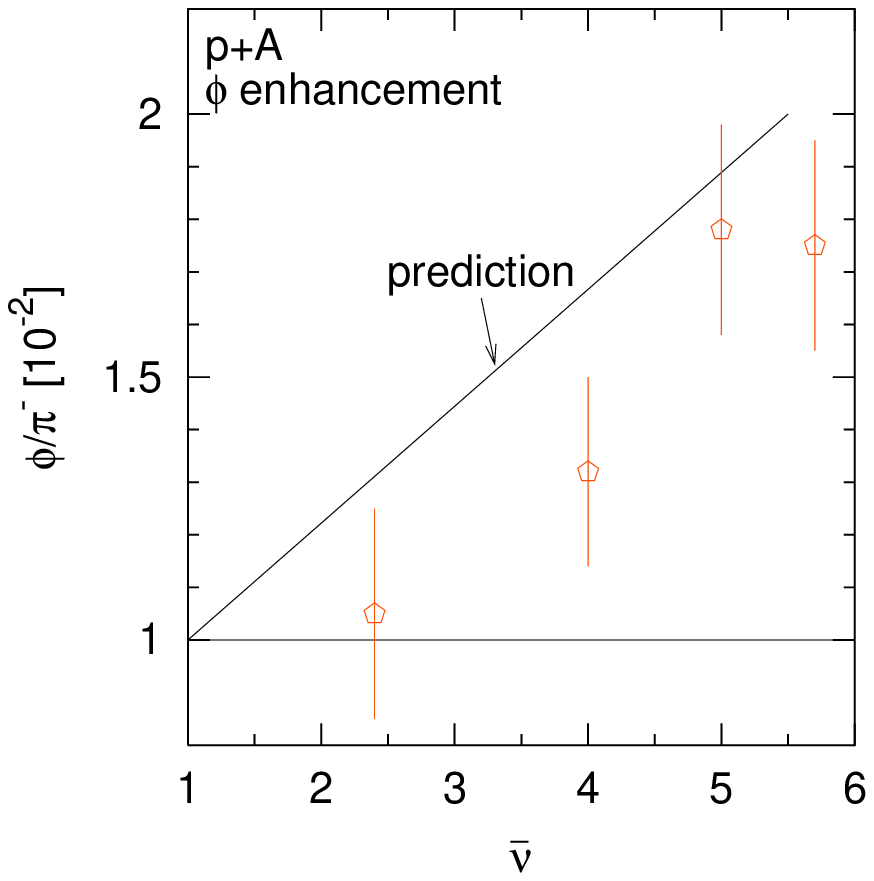}

\caption{{\it a)} Yields of negative pions and $\phi$ in the forward
hemisphere in p+p collision if the fastest proton has longitudinal
momentum $x_{_F}^p$. {\it b)} $\phi$ enhancement in p+A collisions in the
forward hemisphere as function of $\bar{\nu}$. Measured points and
predicted line using p+p correlations are shown. The enhancement for central
Pb+Pb is at 3.}

\label{fig:pred_phi}
\end{figure}

As another example, a similar study can be performed with the $\phi(1020)$
meson that carries hidden strangeness. The ratio $\phi/\pi^-$ in the
forward hemisphere increases if the event has a slower proton
(Fig.~\ref{fig:pred_phi}.a). This leads to a prediction of $\phi$
enhancement in p+A that is in agreement with the data
(Fig.~\ref{fig:pred_phi}.b).

\section{A+A strangeness}

The 10\% most central Pb+Pb collisions have been used. $\Xi^-$ hyperons
decay via the channel $\Xi^- \rightarrow \Lambda \pi^-$ with the
subsequent decay $\Lambda \rightarrow p \pi^-$ following. They are found
by reconstructing the decay vertices starting with the $\Lambda$ decay
vertex. The analyses of $\Lambda(1520)$ and $\phi(1020)$ employ an
alternative method of signal extraction. Here, the signal has been
extracted from the invariant mass spectra after a procedure of mixed event
background subtraction. Corrections for geometrical acceptance, branching
ratio and reconstruction efficiency have been applied.

\subsection{$\Xi^-$ analysis}

\begin{figure}[h]
\epsfxsize160pt
\epsfysize160pt
\figurebox{}{}{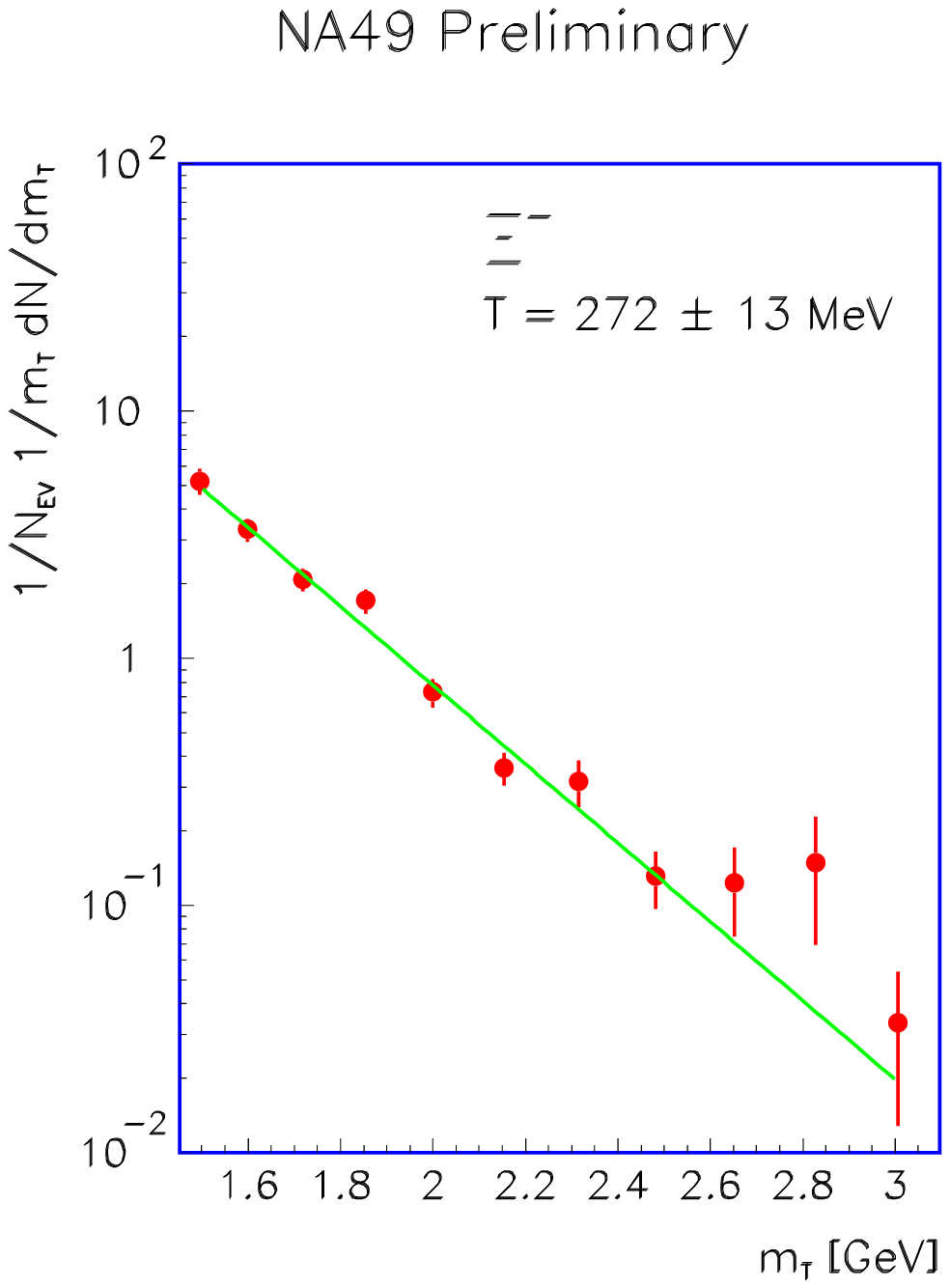}
\figurebox{}{}{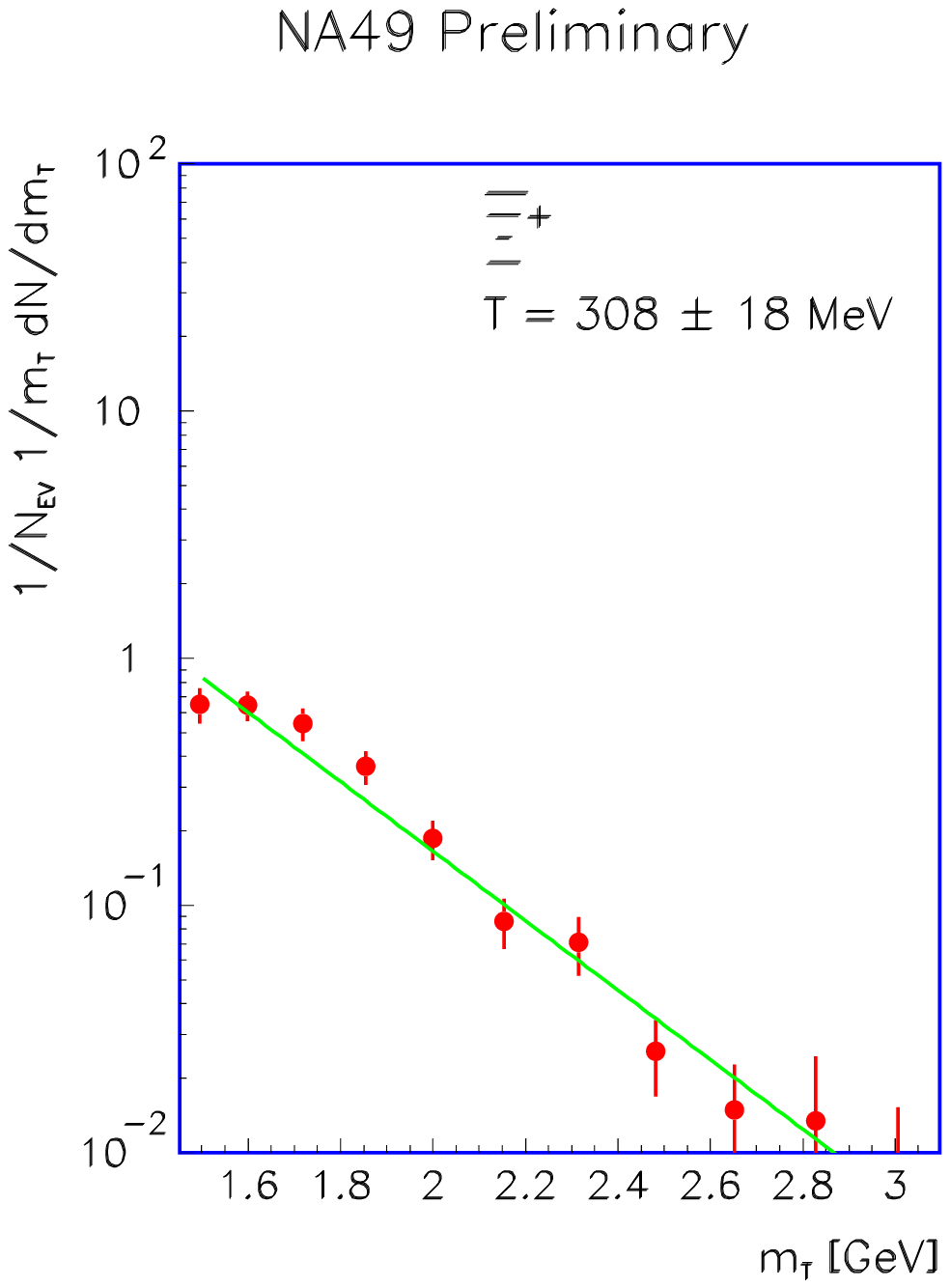}

\caption{Transverse mass distributions for $\Xi^-$ (left) and
$\overline{\Xi}^+$ (right) from central Pb+Pb collisions. Inverse slope
parameters are also given.}

\label{fig:xi_mt}
\end{figure}

\begin{figure}[h]
\epsfxsize110pt
\figurebox{}{}{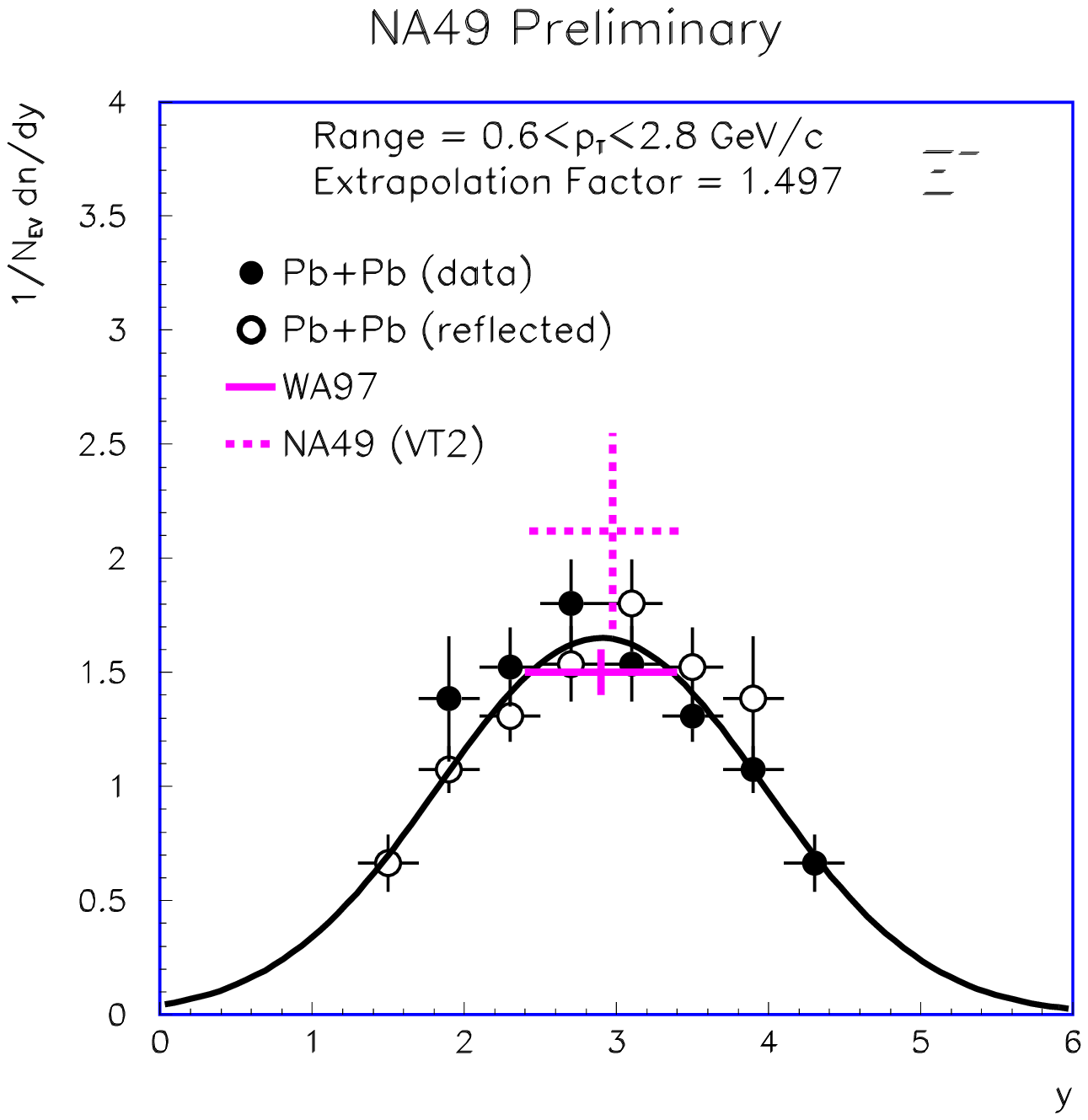}
\figurebox{}{}{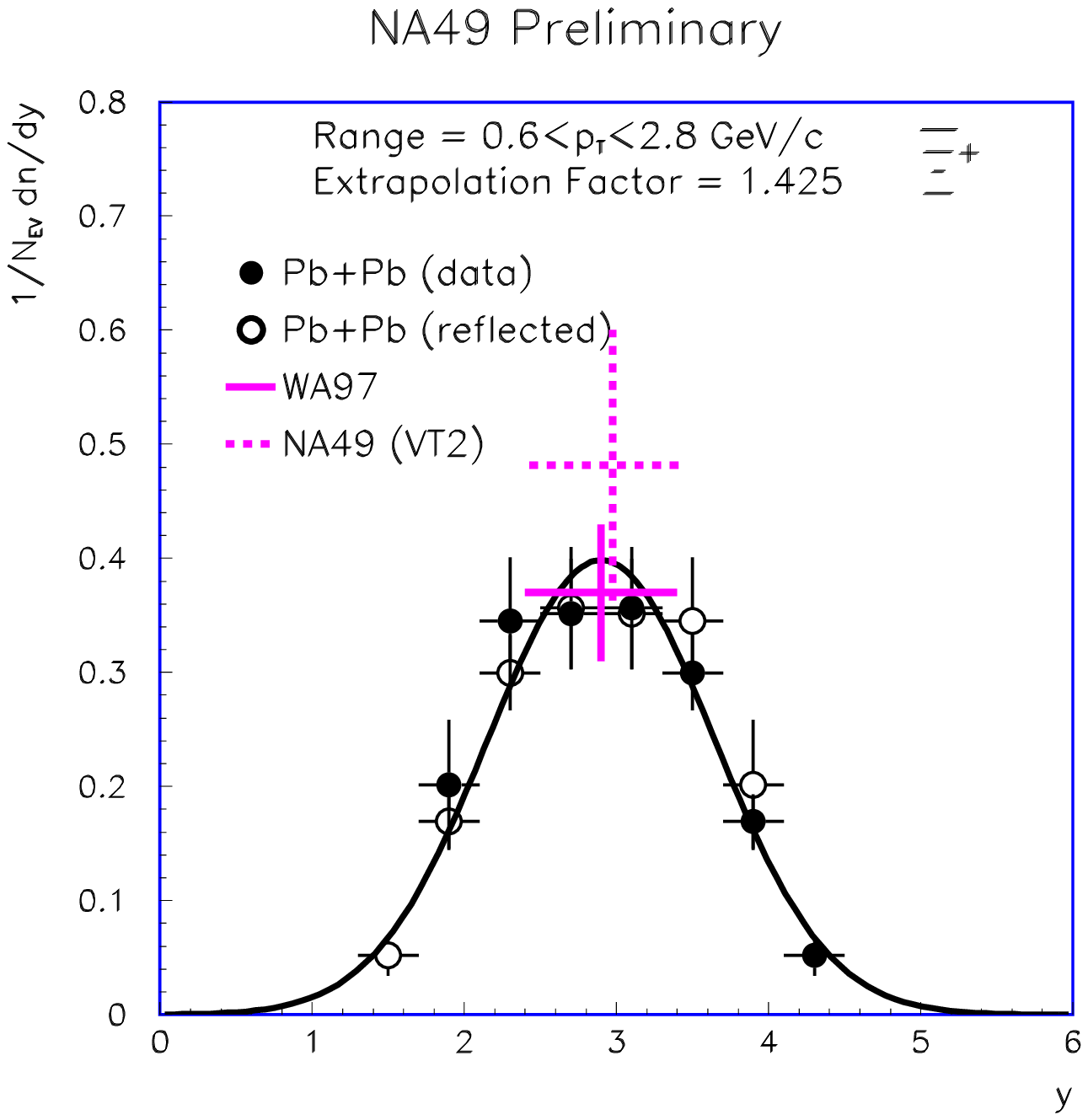}
\figurebox{}{}{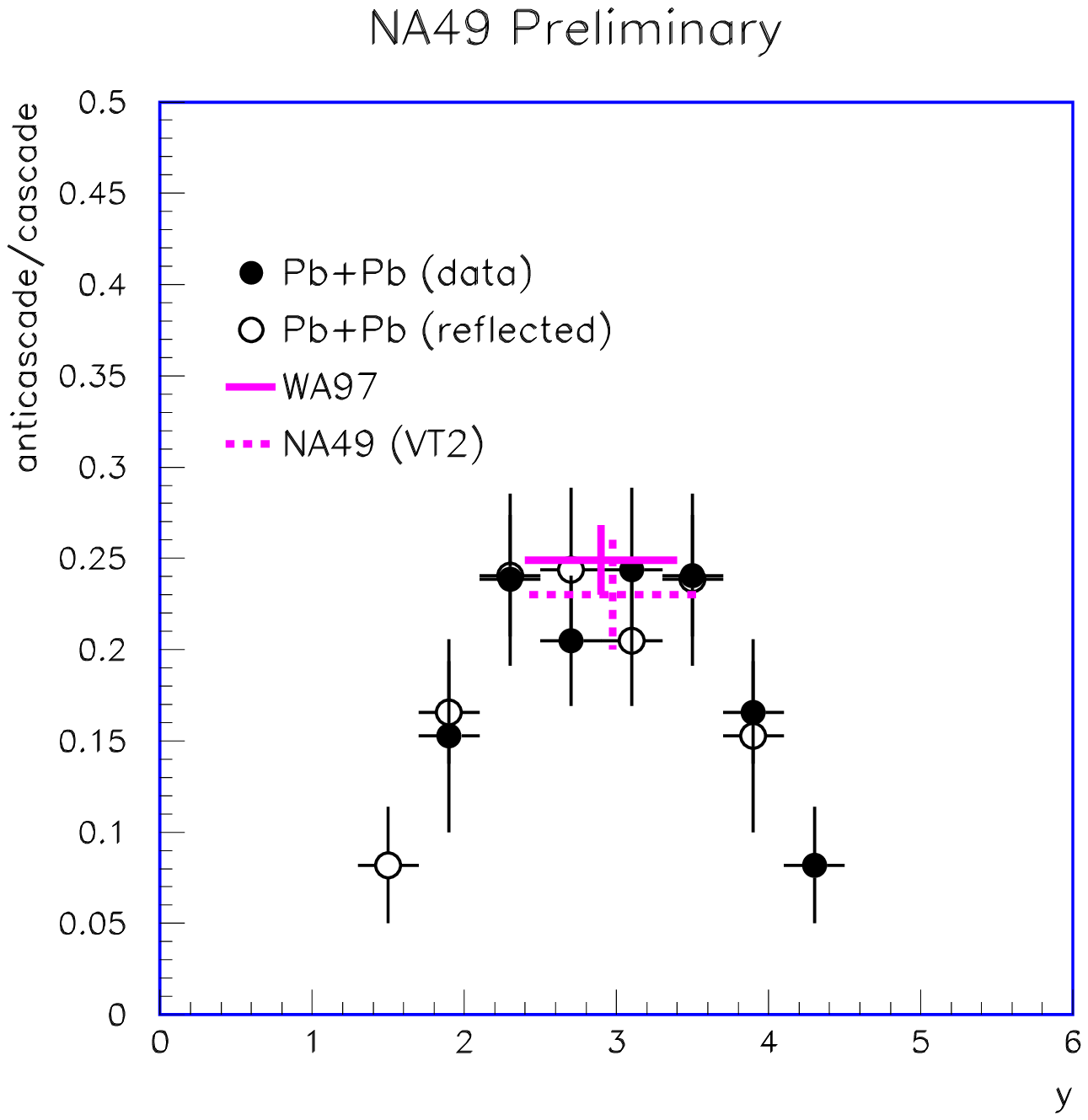}

\caption{Rapidity distribution for $\Xi^-$ (left) and $\overline{\Xi}^+$
(centre) and $\Xi^-/\overline{\Xi}^+$ ratio (right) from Pb+Pb collisions.
Closed circles are measured data points and open circles are reflected
about midrapidity.}

\label{fig:xi_longi}
\end{figure}

Transverse mass (Fig.~\ref{fig:xi_mt}) and rapidity distributions
(Fig.~\ref{fig:xi_longi}) are shown here. Integrating the Gaussian fits
over the full rapidity range gives total yields of $4.42 \pm 0.31$ and
$0.74 \pm 0.04$ particles per event for $\Xi^-$ and $\overline{\Xi}^+$
respectively. The $\Xi^-/\overline{\Xi}^+$ ratio at midrapidity is found
to be $0.22 \pm 0.04$ in good agreement with our previous publication and
other experiment\cite{multi}. The integrated ratio is $0.17 \pm 0.02$.

\subsection{$\Lambda(1520)$ analysis}

Preliminary invariant mass distribution for the $\Lambda(1520) \rightarrow
p+K^-$ channel for p+p and Pb+Pb collisions is obtained. In the case of
p+p collisions the corrected yield amounts to $0.0012 \pm 0.003$ per
event. Scaling the p+p yield by the number of participants to central
Pb+Pb collisions, the expectation would be
$0.012\times(350/2) \approx 2.1\Lambda(1520)$ per event. At the same time
in Pb+Pb only a weak signal is seen which may hint a possible suppression.

\subsection{$\phi(1020)$ analysis}

Invariant mass distributions for the $\phi(1020) \rightarrow K^++K^-$
channel in Pb+Pb and p+p collisions are obtained. No shift in the position
or significant broadening of the mass peak is observed. Inverse slopes of
$305 \pm 15$ and $169 \pm 17$ MeV, integrated yields of $7.6 \pm 1.1$ and
$0.012 \pm 0.0015$ per event are found for central Pb+Pb and inelastic p+p
reactions respectively. This indicates a factor of 3
enhancement\cite{phi}.

\begin{figure}[h]
\epsfxsize320pt
\figurebox{}{}{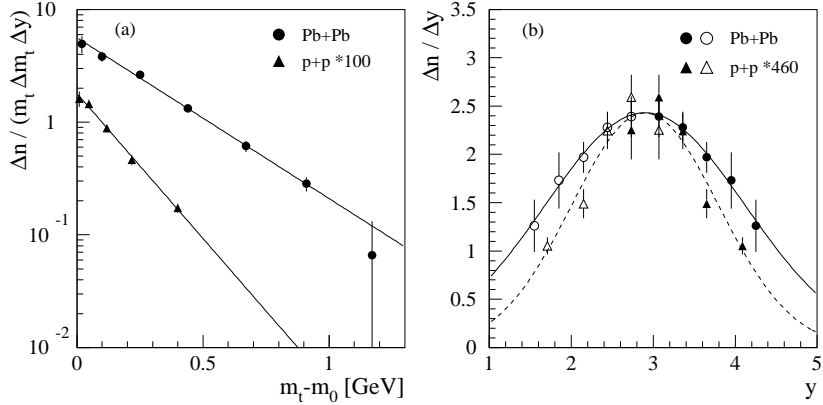}

\caption{$\phi$ mesons in central Pb+Pb and minimum bias p+p collisions:
{\it a)} transverse mass distributions around midrapidity; {\it b)}
rapidity distributions (full symbols represent measured points, open ones
are reflected about midrapidity).}

\label{fig:phi_spec}
\end{figure}

\vspace{-0.1in}
\section{Energy dependence}

In order to explore the energy range between top AGS and top SPS energies
an energy scan has been started. The importance of this study is supported
by a statistical approach that predicts the appearance of a phase
transition to quark-gluon plasma in the early stage of nucleus-nucleus
collisions in this energy range. In the following the results of the
preliminary analysis of 100k reconstructed central Pb+Pb events at 40
GeV$\cdot A$ beam energy are presented.

\begin{figure}[h]
\epsfxsize160pt
\figurebox{}{}{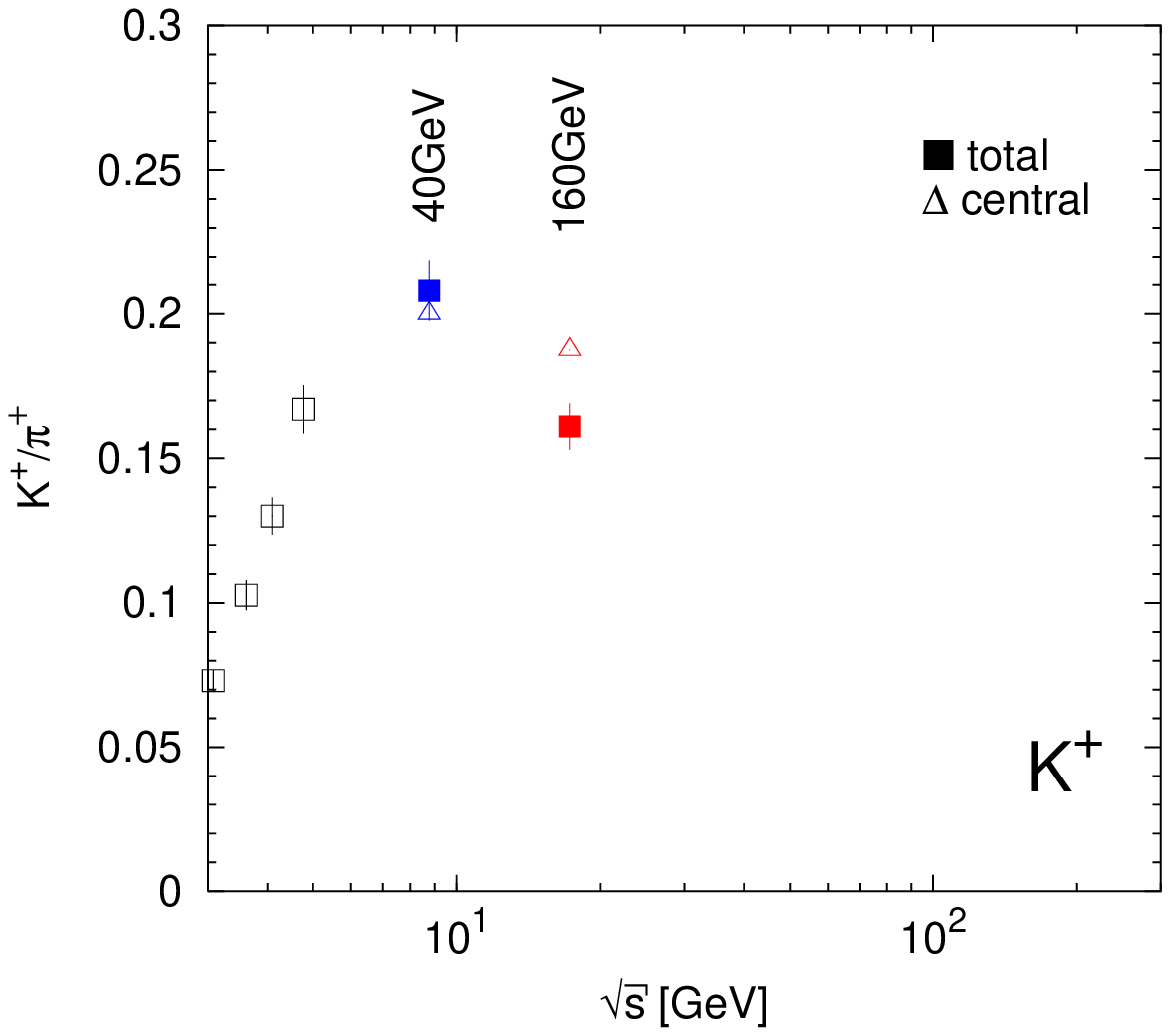}
\figurebox{}{}{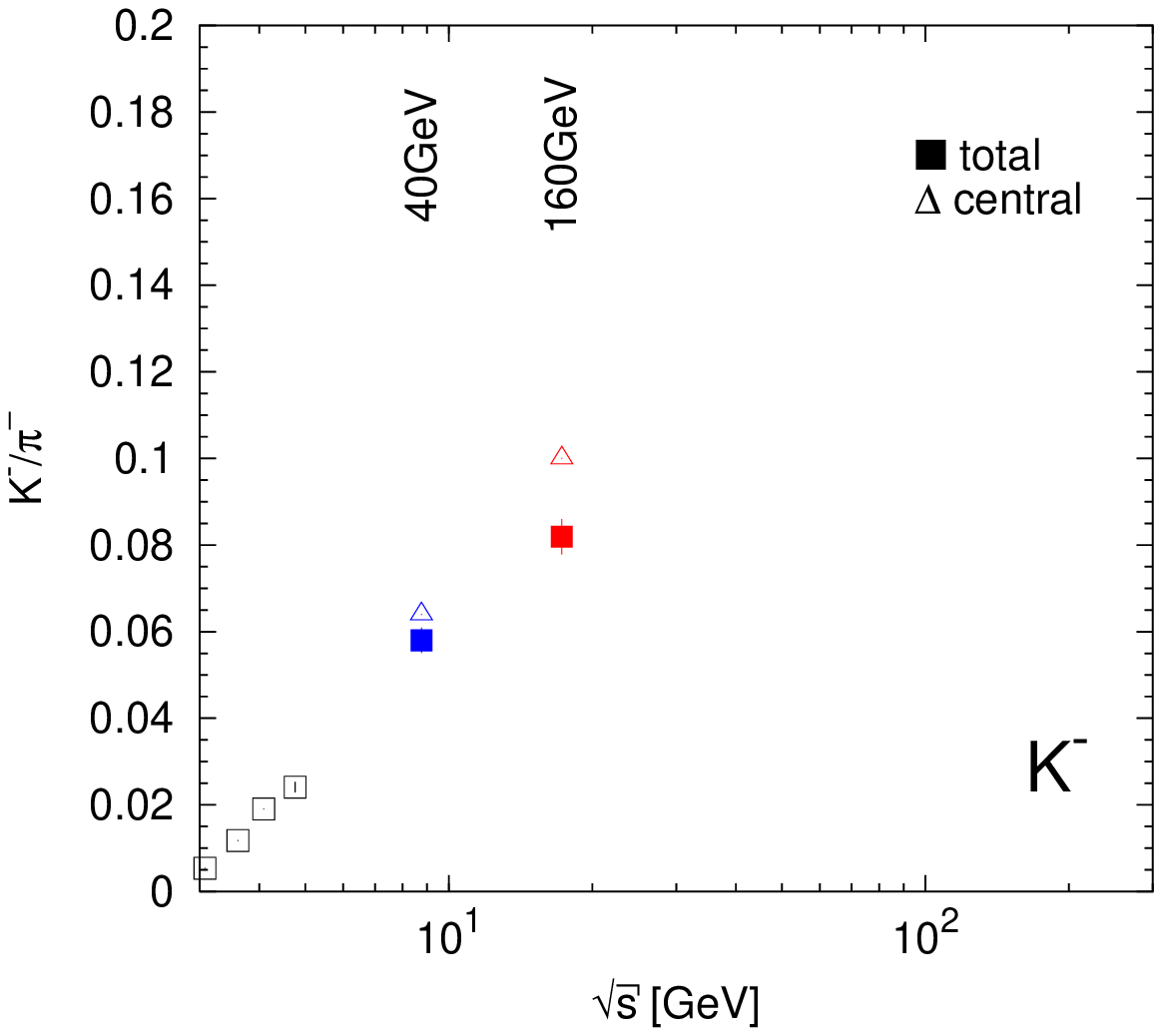}

\caption{Ratios of $K^+/\pi^+$ and $K^-/\pi^-$ yields as function of
collision energy. Closed boxes represent total yields, closed triangles
are for central yields.}

\label{fig:40_ratios}
\end{figure}

\begin{figure}[h]
\epsfxsize160pt
\figurebox{}{}{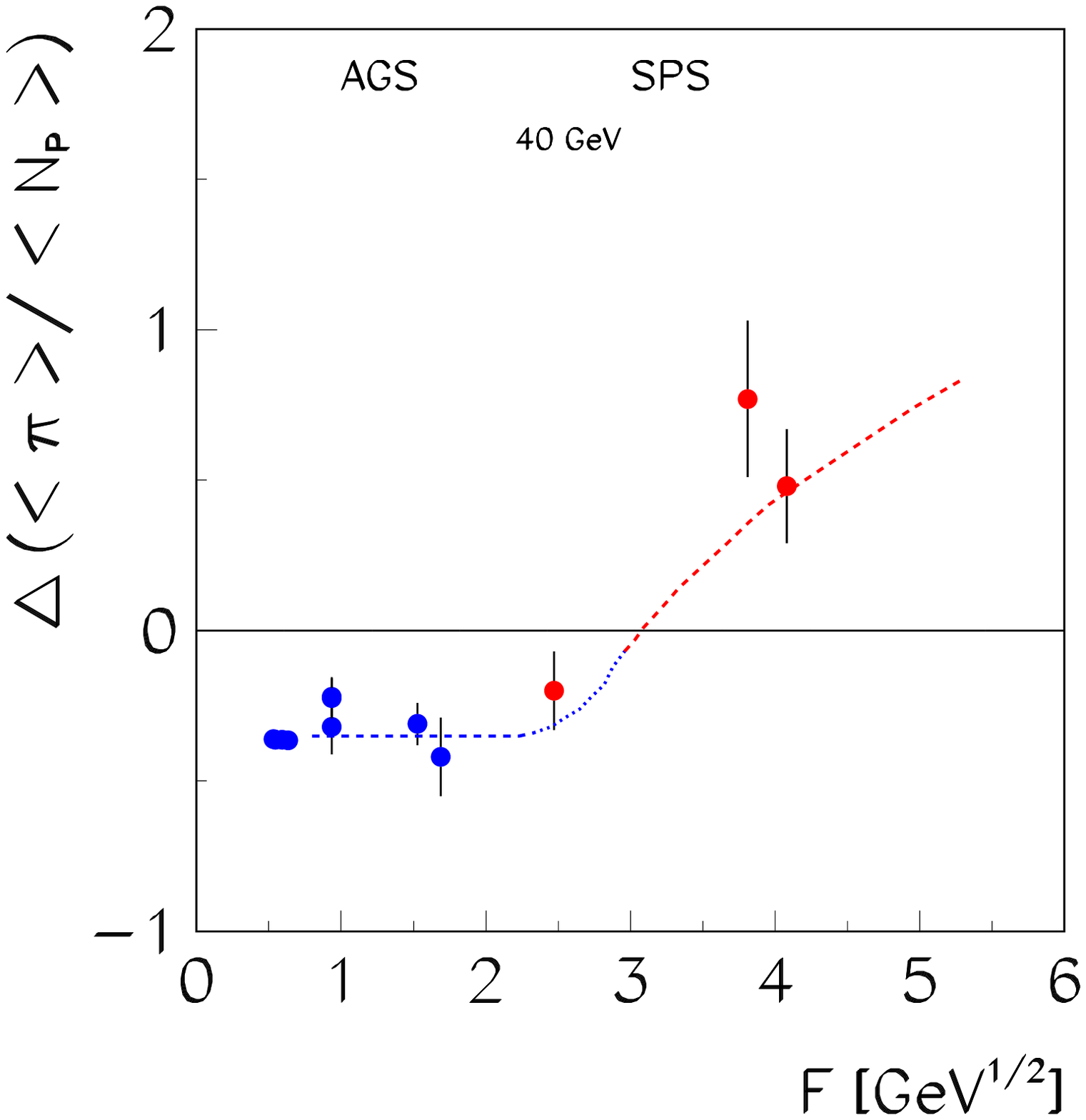}
\figurebox{}{}{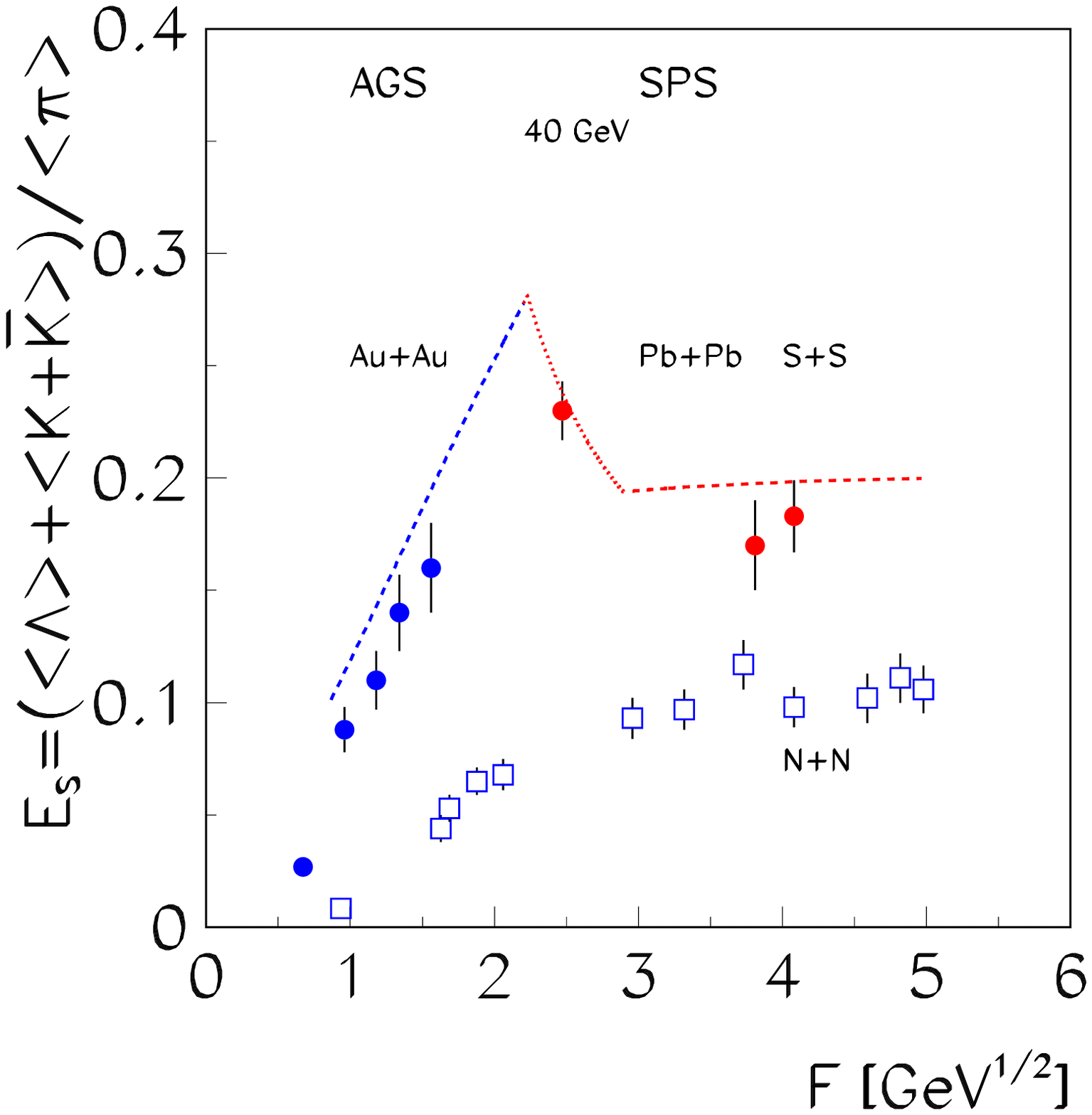}

\caption{The dependence of {\it a)} the difference between pion to
participant ratios -- {\it b)} strangeness to pion ratio -- for central
A+A collisions and N+N interactions on the collision energy expressed by
Fermi's measure. The line indicates the prediction of the statistical
model.}

\label{fig:40_statistical}
\end{figure}

The excitation function of $K^+$ show peak around 40 GeV$\cdot A$
(Fig.~\ref{fig:40_ratios}). The increase of $K^-/\pi^-$ is continuous. The
energy dependence of the ratio of mean pion multiplicity to the number of
participating nucleons (Fig.~\ref{fig:40_statistical}.a) and the
dependence of strangeness production (Fig.~\ref{fig:40_statistical}.b) are
shown with the prediction of the statistical
model of the early state\cite{statistical_model}.

It is observed that the pion yield per participant ratio follows the
established trend of low energy data. The strangeness content of the event
($E_S$) in Pb+Pb collisions at 40 GeV$\cdot A$ is higher by about 30\%
than the corresponding ratios at top AGS and SPS energies. This results
may be the first indication of a non-monotonic energy dependence of the
strangeness to pion ratio.

RQMD shows similar tendency of central yields\footnote{The authors state
that the strangeness increase is mainly due to rescattering of mesons with
baryons. Compare this to the thickness-scaling found in section
\ref{light}.} but fails for total ones\cite{rqmd}.

\begin{figure}[t]
\epsfxsize160pt
\figurebox{}{}{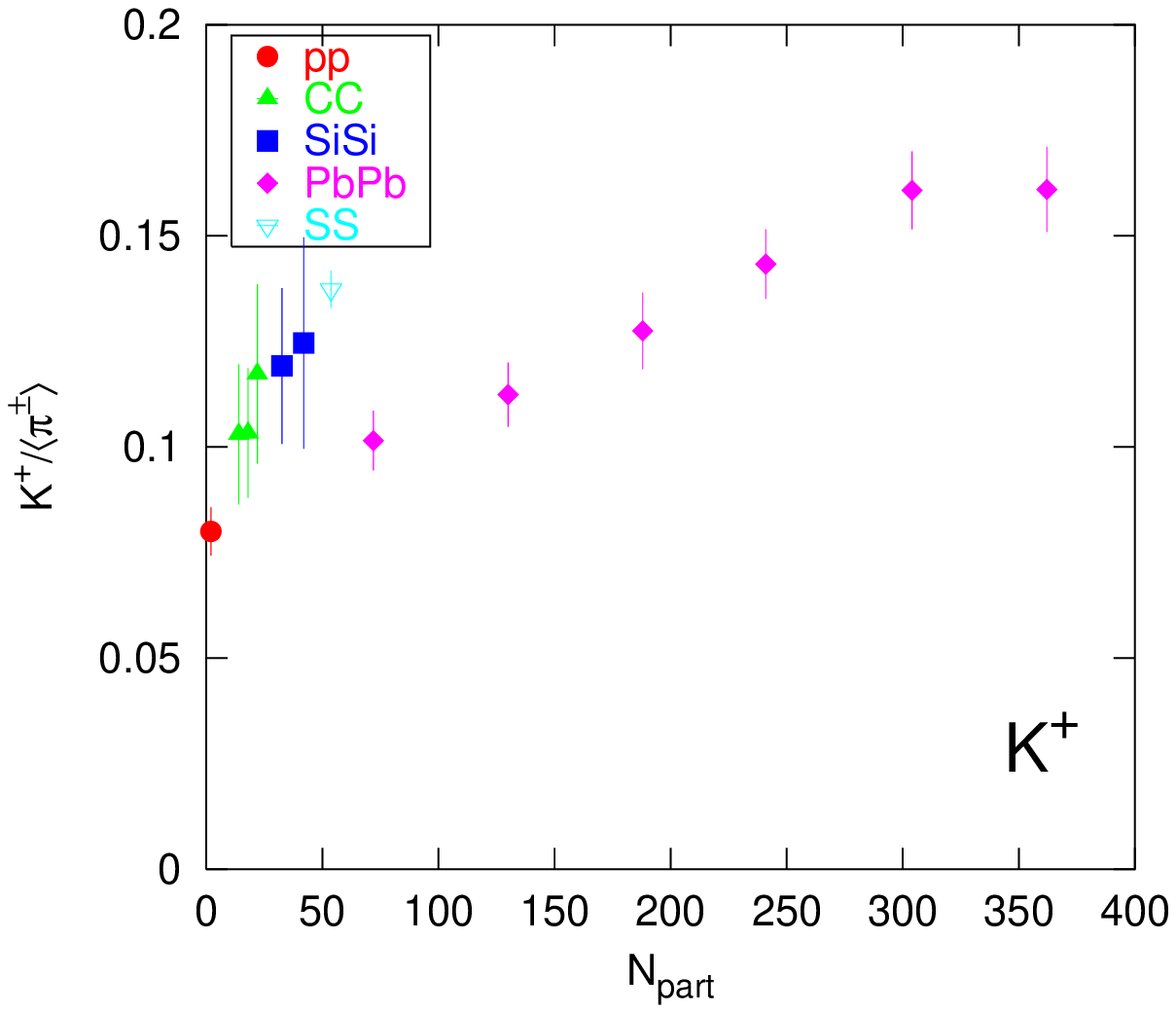}
\figurebox{}{}{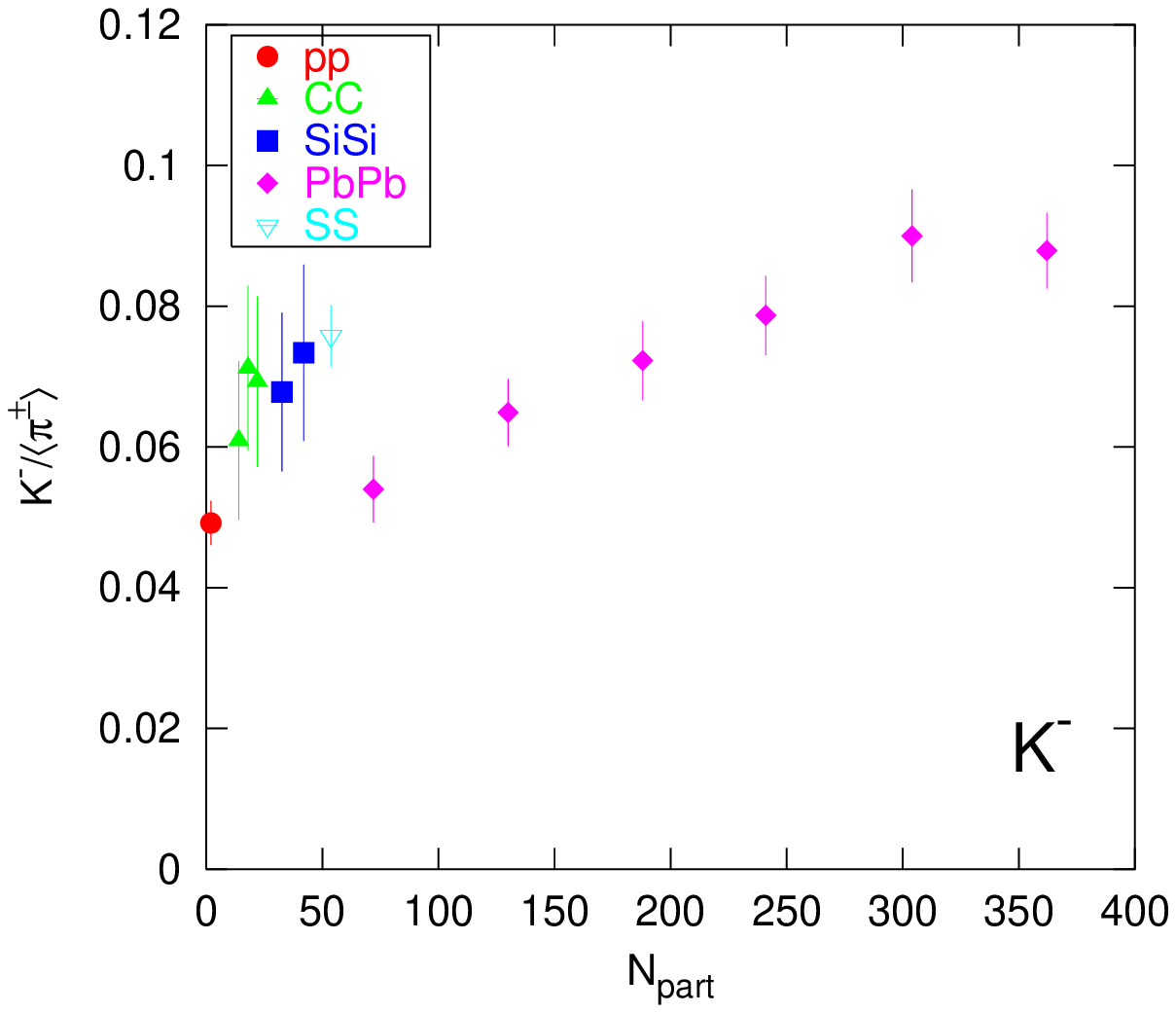}

\caption{Ratios of $K^+/\langle\pi^{\pm}\rangle$ and
$K^-/\langle\pi^{\pm}\rangle$ total yields as function of number of
participants in nuclear collisions.}

\label{fig:npart_scaling}
\end{figure}

\vspace{-0.1in}
\section{System size dependence}

\label{light}

\vspace{-0.05in} Both AGS and SPS experiments observe particle ratios that
show smooth dependence on the centrality of the nucleus-nucleus collision,
that is, on the number of participants ($N_{part}$). On the other hand,
for comparison of interactions of nuclei of different size, $N_{part}$
does not seem to be the right variable, particle ratios do not scale with
the volume of the reaction zone (Fig.~\ref{fig:npart_scaling}, new points
for C+C and Si+Si are from our preliminary analysis).

\begin{floatingfigure}[l]{160pt}
\epsfxsize160pt

\vspace{-0.45in}
\figurebox{}{}{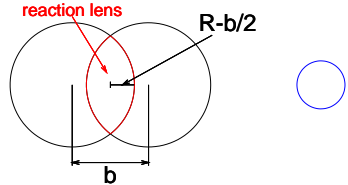}

\vspace{-0.35in}
\caption{Some geometry, the nuclei move inward and outward.}

\label{fig:geom}
\end{floatingfigure}

\begin{figure}[h]
\epsfxsize160pt
\figurebox{}{}{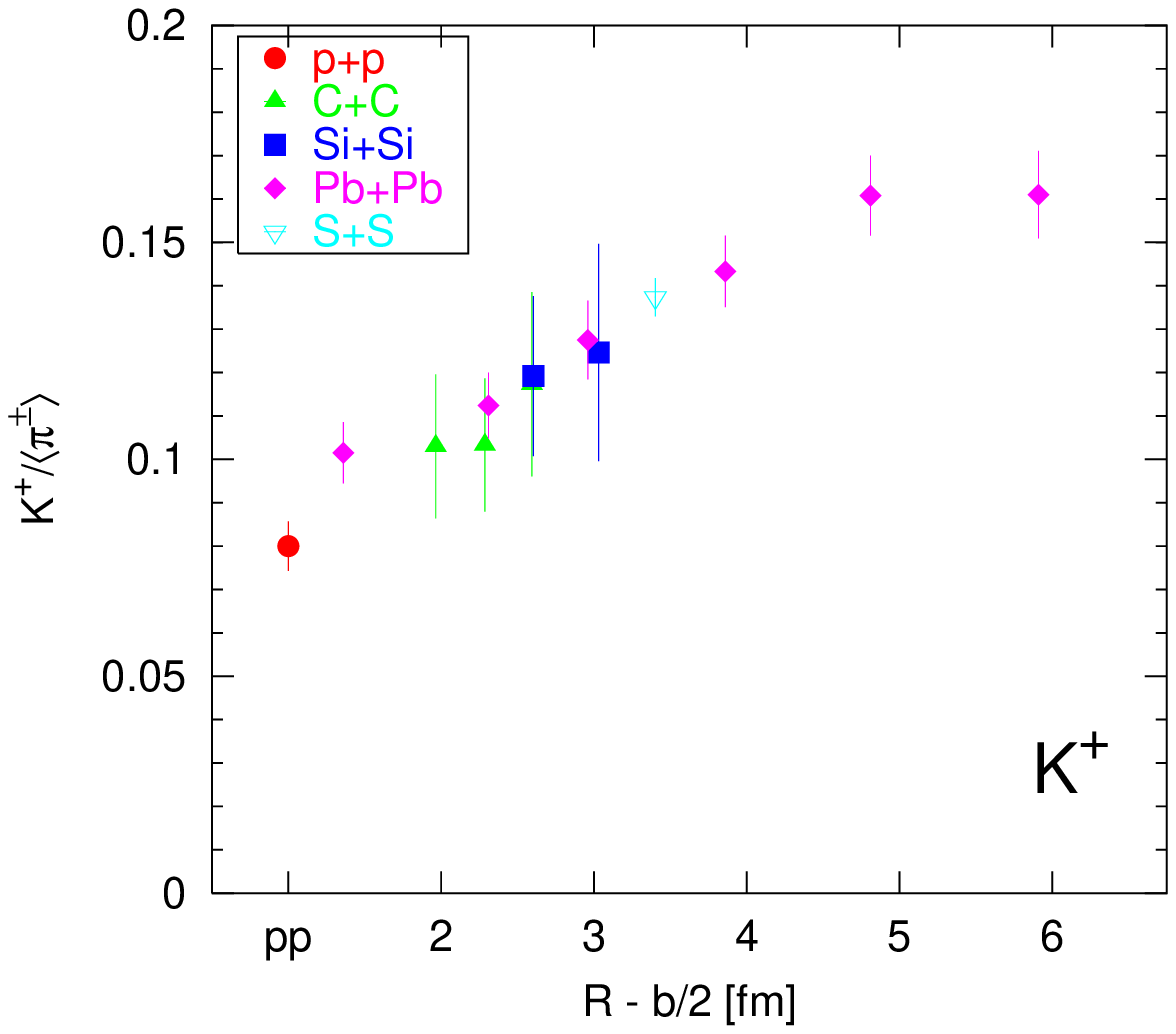}
\figurebox{}{}{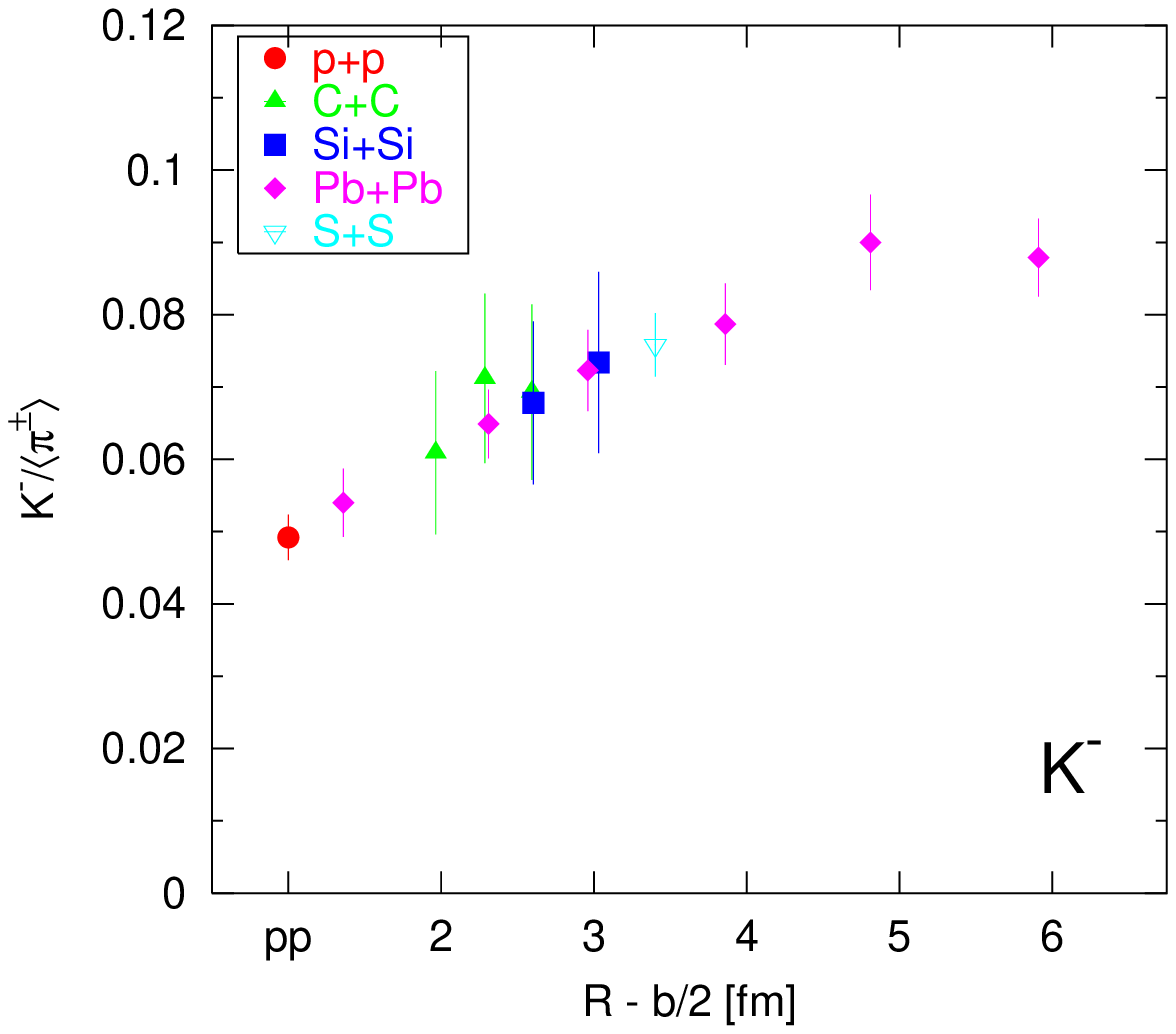}

\caption{Ratios of $K^+/\langle\pi^{\pm}\rangle$ and
$K^-/\langle\pi^{\pm}\rangle$ total yields as function of R-b/2 in
nuclear collisions.}

\label{fig:thickness_scaling}
\end{figure}

Here a new scaling measure, the thickness of the reaction zone is
proposed. Using the cross section and perimeter of the reaction lens, the
volume to surface ratio can be well approximated to be $V/A \approx
\frac{2}{3}(R-b/2)$ which can be regarded as the average escape pathlength
from the reaction volume (Fig.~\ref{fig:geom}\footnote{The central
collision of small nuclei (right) produces as much kaons as the large one
at medium impact (left).}). The impact parameter $b$ can be estimated via
the energy deposited in the zero degree calorimeter. Plotting the above
measures against $R-b/2$ all the four central values (C+C, Si+Si, S+S,
Pb+Pb) and even the Pb+Pb centrality selected points are closely on the
same curve, a line (Fig.~\ref{fig:thickness_scaling}). There are
indications that the increase of transverse momentum for protons is
already present for C+C and Si+Si collisions.

\vspace{-0.35in}
\section{Summary}

\vspace{-0.05in}
The NA49 experiment identifies particles in the forward hemisphere with
controlled centrality. Hadronic reactions with different targets,
projectiles and energies are studied.

Using the internal structure of the p+p interaction, predictions can be
made for p+A and A+A collisions. This is to be confronted with the usual
method of comparing with minimum bias p+p collision.

New results from Pb+Pb collisions explore the characteristics of $\Xi^-$,
$\overline{\Xi}^+$, $\Lambda(1520)$ and $\phi(1020)$ production. Our
energy scan opens up the exciting range of lower energy collisions (40 GeV
$\cdot A$) which has the highest strangeness content observed so far. New
studies with light nuclei help to fill the desert between elementary p+p
interactions and Pb+Pb collisions. Already semi-central C+C collisions
shows increase of strangeness. An interesting scaling with the thickness
of the reaction zone is revealed.

\vspace{-0.1in}


\end{document}